\documentclass[conference]{IEEEtran}
\IEEEoverridecommandlockouts
\usepackage{cite}
\usepackage{amsmath,amssymb,amsfonts}
\usepackage{algorithmic}
\usepackage{graphicx}
\usepackage{textcomp}
\usepackage{changepage}
\usepackage{tabularx}

\usepackage[letterpaper, top=.75in, bottom=1in, left=0.625in, right=0.625in]{geometry}
\usepackage{xcolor}
 \def\BibTeX{{\rm B\kern-.05em{\sc i\kern-.025em b}\kern-.08em
    T\kern-.1667em\lower.7ex\hbox{E}\kern-.125emX}}

\begin{document}

\title{
Pilot Contamination in Massive MIMO Systems: Challenges and Future Prospects
\\

\thanks{This research was partly funded by Palmer Department Chair Endowment Funds at Iowa State University.}
}

\author{\IEEEauthorblockN{ Muhammad Kamran Saeed}
\IEEEauthorblockA{\textit{Department of Electrical and}\\\textit{ Computer Engineering} \\
\textit{Iowa State University}\\
Ames, USA \\
kamran@iastate.edu}
\and

\IEEEauthorblockN{Ashfaq Khokhar}
\IEEEauthorblockA{\textit{Department of Electrical and}\\\textit{ Computer Engineering} \\
\textit{Iowa State University}\\
Ames, USA \\
ashfaq@iastate.edu}
\and

\IEEEauthorblockN{Shakil Ahmed}
\IEEEauthorblockA{\textit{Department of Electrical and}\\\textit{ Computer Engineering} \\
\textit{Iowa State University}\\
Ames, USA \\
shakil@iastate.edu}
}

\maketitle

\begin{abstract}
Massive multiple input multiple output (M-MIMO) technology plays a pivotal role in fifth-generation (5G) and beyond communication systems, offering a wide range of benefits, from increased spectral efficiency (SE) to enhanced energy efficiency and higher reliability. However, these advantages are contingent upon precise channel state information (CSI) availability at the base station (BS). Ensuring precise CSI is challenging due to the constrained size of the coherence interval and the resulting limitations on pilot sequence length. Therefore, reusing pilot sequences in adjacent cells introduces pilot contamination, hindering SE enhancement. This paper reviews recent advancements and addresses research challenges in mitigating pilot contamination and improving channel estimation, categorizing the existing research into three broader categories: pilot assignment schemes, advanced signal processing methods, and advanced channel estimation techniques. Salient representative pilot mitigation/assignment techniques are analyzed and compared in each category. Lastly, possible future research directions are discussed.
\end{abstract}


\section{Introduction}
Massive multiple input multiple output (M-MIMO) has been widely regarded as an essential component of fifth-generation (5G) and beyond communication technologies that enables several benefits spanning from increased spectral efficiency (SE) to enhanced energy efficiency and increased reliability. However, these benefits are contingent upon the presence of accurate channel state information (CSI) at the base station (BS). Timely acquisition of CSI at the BS is crucial for maximizing network throughput. However, acquiring CSI is often challenging, primarily due to small coherence intervals. 

\par Coherence interval refers to a time duration wherein the channel response remains frequency-flat and time-invariant. Generally, the coherence interval is shorter in scenarios characterized by rapid changes in the wireless channel, such as high mobility, narrow bandwidth, change in the propagation environment, etc. This limitation results in reusing the same pilot sequences in adjacent cells, preventing orthogonal distribution to all users. This introduces coherent interference into channel estimation, as multiple devices transmit the same pilot sequence, making it challenging for the BS to distinguish each device's pilot sequences. This phenomenon is called pilot contamination. Pilot contamination is considered a critical limiting factor in enhancing SE \cite{intro1}.
 
 \par  Though a comprehensive survey article \cite{intro2} on mitigating pilot contamination was published in 2015, most discussed schemes are now outdated. Therefore, this paper summarizes the most recent developments in addressing pilot contamination, driven by rapid research progress. It categorizes the pilot decontamination schemes into three broader categories: pilot assignment schemes, advanced signal processing methods, and advanced channel estimation techniques. Pilot assignment schemes intelligently allocate pilot sequences to minimize interference and ultimately maximize SE \cite{spa1}, \cite{gc1}, \cite{pr1}. Advanced signal processing strategies introduce innovative pilot transmission and signal processing techniques that effectively mitigate interference and enhance throughput \cite{si1}, \cite{si2}, \cite{rs2}. Pilot decontamination through advanced channel estimation techniques, such as deep learning, aids in mitigating the mean square error (MSE) of the channel estimate, eventually improving CSI and enhancing SE. This paper analyzes these techniques, compares and contrasts using key features and performance indicators.  

\section{Channel State Information}
The time-frequency is segmented into coherence intervals/blocks, $\tau_c$, during which channel response remains time-invariant and frequency-flat. Within each block, a designated interval, $\tau_{\rho}$, is reserved for transmitting predefined pilot signals during uplink transmission. Pilot signals with predefined patterns are transmitted, enabling the BS to compare the transmitted and received values, as the received pilot signal may contain valuable information, such as multi-path fading, interference, and noise, as it transverses the wireless channel.

\subsection{Pilot Contamination}
In M-MIMO systems, the limited coherence block length constrains the availability of lengthy pilot signals. This limitation restricts the distribution of unique/orthogonal pilot sequences among all users, leading to the necessity of reusing the same pilot sequences in the adjacent cells. The uniformity of pilot signals originating from devices in various cells creates a challenge for the BS to distinguish between the desired pilot signal (from the serving cell) and the interfering pilot signals (from neighboring cells). This phenomenon, known as inter-cell pilot contamination, and is recognized as a limiting factor in enhancing SE. Fig. 1, shows a $L$ cells M-MIMO network with $K$ users per cell. The $k_{th}$ user, $U_k$ in different cells are utilizing the same pilot sequence as of $k_{th}$ user, $U_k$ in $i_{th}$ cell, causing a detrimental impact on $k_{th}$ user CSI in the $i_{th}$ cell.  
\vspace{-0.3cm}
\begin{figure}[htp]
    \centering
    \includegraphics[width=8.5cm]{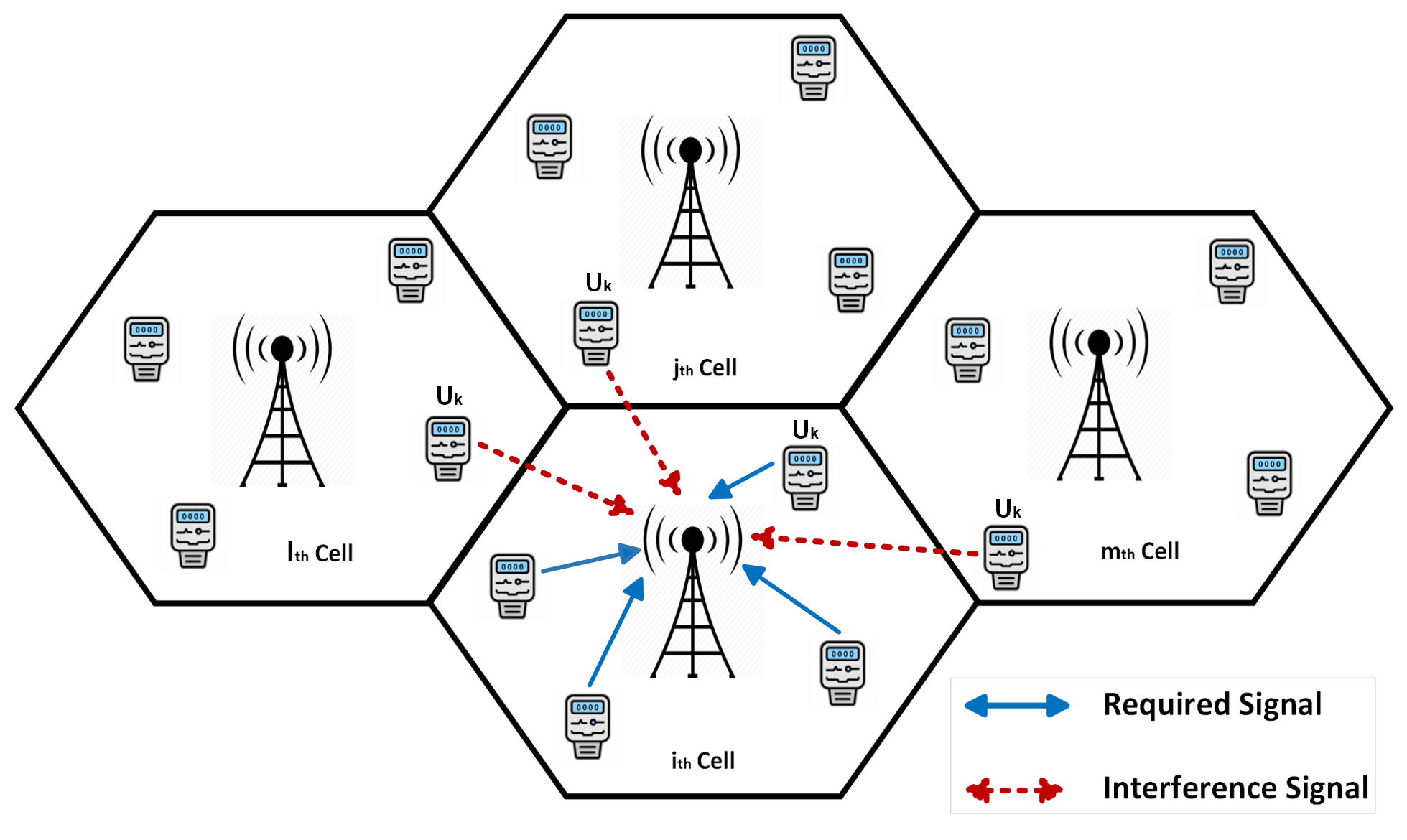}
        \vspace{-0.2cm}
    \caption{Desired and interference signals}
    \label{fig:3}
\end{figure}
\vspace{-0.1cm}
\vspace{-0.3cm}

\section{Mitigating Pilot Contamination through Pilot Assignment Schemes
}
This section explores representative pilot assignment strategies to mitigate pilot contamination in M-MIMO systems.

\subsection{Smart Pilot Assignment Schemes}
Smart pilot assignment schemes intelligently allocate pilot signals to users with the aim of minimizing pilot contamination to improve SE. Zhu et al. in \cite{spa1} proposed a pilot scheme that optimized pilot assignment within a target cell while addressing interference from users using the same pilot signals in neighboring cells, aiming to maximize the minimum Signal-to-Interference-plus-Noise-Ratio (SINR). Subsequently, the pilot signal with the least interference from neighboring cells was assigned to the target cell's user with the worst channel quality in a sequential manner until all users were allocated.

Zhu et al.'s scheme was simple but had high complexity and relied on unrealistic assumptions of uncorrelated Rayleigh fading, limitations that were addressed by \cite{spa2}. This paper aimed to maximize the minimum weighted sum SE for each user in the target cell. A low-complexity heuristic algorithm prioritized users with the worst-weighted SE. This algorithm iteratively assigned low-interference pilots to users with the worst channel conditions across all cells until all pilots were allocated. In conclusion, smart pilot schemes are simple and yield favorable results by intelligently allocating pilot signals across cells.

\subsection{Graph Coloring-based Assignment Schemes}
Graph coloring, a technique in graph theory, involves assigning colors to vertices (nodes/users) such that adjacent vertices have distinct colors, aiming to use the fewest colors possible. This method offers significant advantages in pilot contamination scenarios: it optimizes the efficient use of limited pilot signal resources by minimizing the number of colors (pilots) used and maximizes the separation between vertices based on defined criteria, such as interference graphs.

Liu et al. \cite{gc1} proposed a graph coloring-based pilot assignment scheme leveraging interference graphs to minimize interference between users by assigning distinct pilots to connected nodes. To efficiently use limited orthogonal pilot sequences, they employed an iterative method to establish a fair pilot reuse policy, preventing the overuse of a few pilots. Numerical results validated the scheme's effectiveness in achieving higher SE while maintaining low complexity. However, a drawback of this paper lies in its consideration of interference as unidirectional, while the real scenarios involve bidirectional interference.

Zeng et al. \cite{gc2} introduced a pilot assignment scheme based on the weighted graph framework that accounts for bidirectional interference. The authors formulated the pilot assignment problem as an optimization challenge, aiming to maximize the total throughput across all users. Moreover, they addressed the pilot assignment problem by employing the max k-cut problem. Simulation results demonstrated the proposed scheme's significant improvement in SE. However, a prominent limitation is the high computational complexity of the max-cut problem with an increasing number of users, thus hindering scalability.

Saeed et al. \cite{gc3} addressed challenges from Zeng et al. \cite{gc2} by introducing an intelligent user scheduling strategy to optimize M-MIMO systems, effectively tackling pilot contamination and scalability issues. To address scalability concerns, they introduced a novel approach: rather than assigning orthogonal pilot sequences to individual devices, they allocated them to clusters of devices \cite{gc4}. This enabled multiple devices to efficiently share the same pilot for periodic data transmission. Furthermore, they utilized a graph coloring framework and leveraged the integer linear programming to address pilot assignment challenges effectively. The proposed scheme substantially reduced complexity and enabled scalability for M-MIMO systems.

\subsection{Soft Pilot Reuse-Based Assignment Schemes}
Soft pilot reuse assignment schemes categorize cell users into cell-center and cell-edge users. Cell-center users are those in close proximity to the BS, typically with strong SINR, where interference minimally impacts their performance. Therefore, pilot sequences allocated to these users can be reused in every other cell. Conversely, cell-edge users are located farthest from the serving BS and experience poorer channel conditions, where interference significantly degrades their performance. Hence, orthogonal pilot sequences are assigned uniquely within each cell's edge user and are not reused in neighboring cells to mitigate interference.

Zhu et al. \cite{pr1} proposed a soft pilot reuse scheme with a multi-cell block diagonalization precoding scheme for an M-MIMO system. They categorized users into two groups based on the interference received from the neighboring cells. By slightly increasing the number of pilot signals compared to traditional approaches, their proposed scheme effectively eliminates pilot contamination experienced by edge users who would otherwise suffer from severe pilot contamination in conventional schemes. Simulation results validated the scheme's ability to improve the QoS for edge users. However, a notable drawback of this paper is its increased pilot overhead, ultimately impacting SE.

To overcome the challenge of \cite{pr1}, Li et al. \cite{pr2} analyzed the downlink performance of the M-MIMO system considering both soft pilot and soft frequency reuse techniques. In soft frequency reuse, the total frequency bands are partitioned into three sub-bands denoted as $F_1$, $F_2$, and $F_3$. Here, neighboring cells' edge users must use different frequency bands while the remaining bands are uniformly allocated to the cell-center group. Similarly, the orthogonal pilot sequences are organized into four sets: $P_1$, $P_2$, $P_3$, and $P_4$. Notably, the pilots for cell-edge users in the target cell were orthogonal to neighboring cells' edge users. Where the cell-center group employed the same set of pilot sequences across all cells. Numerical results confirmed the efficacy of the scheme in achieving higher SE.

\subsection{Angle of Arrival-Based Assignment Schemes}
Much of the existing research on pilot assignments primarily centered on large-scale fading coefficients, which are significantly dependent on the distance between users and the BS. In \cite{aoa1}, it was proposed to reconsider pilot assignment solely based on large-scale fading coefficients by introducing the angle of arrival of pilot signals as a consideration, aiming for contamination-free estimation and improved spectral efficiency.

Shahabi et al. \cite{aoa1} presented an angle of arrival-based pilot allocation strategy that effectively combined angular and static channel information. This paper introduced an optimization problem and claimed that it had significantly lower complexity when compared to methods like smart pilot assignment \cite{spa1}. To achieve this, the pilot assignment problem was formulated based on the multi-path channel model and the utilization of users' angle of arrival distributions. The proposed scheme achieved faster convergence and lower complexity. However, the limitation of this paper led to a decrease in uplink performance compared to conventional methods.

To address this challenge, Omid et al. \cite{aoa2} proposed an angle of arrival-based pilot allocation scheme using deep reinforcement learning. Firstly, they designed a cost function that assessed pilot contamination based on user locations and channel quality. Furthermore, they defined sets of states, actions, and reward functions based on channel characteristics; the agent learned a pilot assignment policy that adapted to channel changes while minimizing the cost function. Numerical results claimed their method could efficiently track various channel scenarios and select pilot assignments close to those obtained through exhaustive searches.

\section{Mitigating Pilot Contamination with Advanced Signal Processing}
This section reviews techniques aimed at enhancing SE using the latest signal processing advancements.
\subsection{Superimposed Pilots Based Interference Mitigation}
In conventional time-multiplexed pilots, a portion of the coherence block is reserved solely for transmitting pilot symbols, which is essential for extracting CSI. However, this exclusive allocation of resources can limit the overall capacity available for transmitting data. In contrast, the superimposed pilot scheme eliminates the need for a dedicated space for pilot transmission by embedding pilot information within data symbols. This allows the receiver to estimate the channel and extract data using the same symbols. This scheme optimizes resource utilization and reduces extra signaling overhead.

Lago et al. \cite{si1} proposed a pilot decontamination approach for M-MIMO networks by leveraging superimposed and time-multiplexed pilots, allowing users to utilize both for channel estimation. The key concept was to utilize the superimposed pilots to address the channel estimation errors obtained through time-multiplexed pilots. Moreover, the proposed scheme outperformed conventional methods, even in a universal pilot reuse scenario, and eliminated the need for inter-BS exchange of user channel statistics, reducing communication overhead. Numerical results showed significant enhancement compared to approaches relying solely on time-multiplexed or superimposed pilots for channel estimation. However, a prominent limitation persisted due to the continued utilization of time-multiplexed pilots, causing substantial communication overhead.

Shafin et al. \cite{si2} overcame the challenge and introduced a superimposed pilot framework for M-MIMO systems, significantly enhancing its practical application. They developed an uplink direction of arrival estimation method, effectively reducing uplink overhead. Additionally, their work comprehensively characterized network throughput in a superimposed pilot-based massive full-dimension MIMO system and validated analytical results through extensive simulations, and identified novel design insights, all within a versatile framework.

\subsection{ Rate Splitting Strategy for Pilot Decontamination}
Rate-splitting multiple access (RSMA) is an emerging strategy for interference mitigation in wireless networks,  positioning itself as a promising paradigm for 6G physical layer transmission. RSMA divides each user's message into common and private components, with common messages decoded collectively and private messages decoded individually by the user equipment, followed by transmission and superimposition of the common message onto the private one. Notably, RSMA offers robustness to imperfect CSI, showcasing superior throughput and energy efficiency in multi-user MIMO networks.

In \cite{rs3}, the authors proposed RSMA integration in a single-cell M-MIMO network, deriving throughput expressions and establishing capacity lower bounds based on channel hardening for both common and private data messages. The authors introduced three power allocation algorithms to maximize various network utilities for RSMA and conducted extensive simulations comparing RSMA with No RSMA under diverse conditions. Results showed RSMA's superiority in mitigating pilot contamination and offering improved SE, especially with imperfect CSI at the transmitter. However, a notable drawback was that this scheme was designed for cell-based M-MIMO and lacked applicability to emerging cell-free M-MIMO technology.

Therefore, Mishra et al. \cite{rs1} proposed an RSMA approach for cell-free machine-type communication with random access, where all active users shared the same pilot for channel estimation. They incorporated RSMA with conventional conjugate beamforming for private data streams to mitigate downlink interference. Initially, the authors first derived achievable throughput expressions for common and private data messages. Then, they used these expressions to devise a heuristic precoding technique for the common message and introduced a novel power control algorithm based on successive convex approximation. They provided numerical evidence to show the effectiveness of RSMA techniques in mitigating pilot contamination.

Cell-free M-MIMO setups in \cite{rs1} assumed perfect synchronization, which was impractical for distributed networks due to inevitable signal delays, thus hindering coherent transmission. Therefore, in \cite{rs2}, the authors examined the data transmission performance of a cell-free M-MIMO system, considering both coherent and non-coherent scenarios. Moreover, they derived a closed-form expression for the sum SE of RSMA-supported cell-free M-MIMO systems, addressing asynchronous reception and optimizing power allocation between common and private messages. Finally, a robust precoding design for common messages was proposed to improve performance, especially in mitigating the impact of asynchronous reception. Numerical results claimed the efficacy of their method in addressing asynchronous reception.

\section{Pilot Decontamination using Advanced Channel Estimation Methods}
Deep Learning (DL) offers a promising solution for complex wireless communication challenges, boosting the system's performance with low computational complexity in tasks like resource allocation, channel decoding, and channel estimation. Xu et al. \cite{dl1} proposed a DL-based pilot design approach for M-MIMO systems, addressing the challenge of optimizing power allocation for each user's pilot sequence to minimize the cumulative mean square error in channel estimation. Their solution involved a deep neural network that mapped the input data (large-scale fading coefficients) to output data (pilot power allocation vector). Simulation results claimed that the proposed scheme surpassed conventional pilot decontamination methods while having lower computational complexity. One limitation of the paper was its reliance on unsupervised learning for pilot power allocation, which may not guarantee optimal solutions.

Therefore, Hirose et al. \cite{dl2} presented two innovative channel estimation techniques: one employed a neural network (NN), and the other utilized a convolutional neural network (CNN). The NN-based approach relied on fully connected layers, focusing on extracting spatial information from the least squares estimated channel. In contrast, the CNN took advantage of the user's spatial correlation using sliding convolutional filters. Their results demonstrated that CNN-based estimation outperformed the NN-based alternative in accuracy, despite slightly longer training times for datasets. However, the paper might have lacked a thorough investigation into the scalability of the proposed DL-based channel estimation method for M-MIMO systems with a large number of antennas or users.

In \cite{dl3}, the authors proposed a low-complexity channel estimation technique for M-MIMO systems by using a deep neural network (DNN) for denoising before the conventional least squares operation, enhancing performance through a tailored DNN architecture. To maintain low complexity, they reconfigured this architecture's input and output layers and optimized the number of parameters, substantially reducing it from millions to just a few thousand parameters. While the proposed scheme demonstrated favorable outcomes, the paper might not have fully explored the computational complexity and resource requirements associated with deploying untrained DNNs in real-time M-MIMO systems, which could have been crucial factors for practical implementation.

Lim et al. \cite{dl4} proposed a method to mitigate pilot contamination in wireless systems using DL. They introduced an NN-based pilot design to minimize mean square error, augmented with unsupervised learning to remove the necessity of using pilot samples during the uplink training phase. Additionally, they utilized deep residual learning for the channel estimator to reduce distortion noise caused by pilot contamination and to reconstruct the original channel. Finally, they introduced a novel pilot design and channel estimator using transfer learning, which outperformed conventional linear estimators, even in the presence of hardware impairments, showing its capability to mitigate interference without prior knowledge.
\begin{adjustwidth}{-1in}{-1in}
\begin{table*}
\caption{Comparative Analysis of the Proposed Schemes.}
    \centering
    \tiny
   \begin{tabular}{|p{1cm}|p{3cm} |p{2cm} |p{1.85cm} |p{2.15cm} |p{2.75cm} |p{1cm} |p{1cm} |} \hline 
 
         \textbf{Ref. No.}&  \textbf{Major Contribution}&  \textbf{System Considered}&  \textbf{Channel Assumption}&  \textbf{Combining/ Precoding Scheme}&  \textbf{Computational Complexity}&  \textbf{Pilot Overhead}& \textbf{Enable Scalability}\\ \hline 
 \multicolumn{8}{|c|}{\textbf{Mitigating Pilot Contamination through Pilot Assignment Schemes}}\\  \hline  
       Zhu et al.  \cite{spa1}  &  Maximize the minimum  SINR via pilot assignment&    Uplink Multicell M-MIMO &  Uncorrelated Rayleigh fading&  MF combining &  $O(K \log K)$& Not addressed & Not addressed
\\ \hline  
      Nguyen et al. \cite{spa2}  &  Maximizing SE through optimal pilot assignment& Uplink-Downlink Multicell M-MIMO &  Correlated Rayleigh fading&  MR combining &  $O(\nu N_1 L^2 K M^3 + N_2^\alpha \max\{2 L^3 K^3, F_1\})$& Not addressed & Not addressed
\\ \hline  
      Liu et al.  \cite{gc1} &  Mitigate pilot contamination via pilot assignment&  Uplink cell-free M-MIMO &  Uncorrelated Rayleigh fading&  MR combining& $ O(K(K + 2M) + KM \log(2M))$& Not addressed & Not addressed
\\ \hline  
      Zeng et al.  \cite{gc2}  &  Optimizing pilot assignment using Weighted Graphs&    Uplink-Downlink cell-free M-MIMO &  Uncorrelated Rayleigh fading&  MMSE combining&  $O\left(\frac{K^2}{2} + \frac{K}{2} + \tau\right)$& Not addressed & Not addressed
\\ \hline  
       Saeed et al. \cite{gc3} &  Mitigating pilot contamination and enabling scalability in M-MIMO&    Uplink Multicell M-MIMO &  Correlated Rayleigh fading&  MMSE combining& $ O(K(K-C))$&  Decreased& Yes
\\ \hline  
       Zhu et al. \cite{pr1}&  QoS improvement through soft pilot reuse&    Uplink-Downlink multicell M-MIMO &  Uncorrelated Rayleigh fading& Multicell block diagonalization precoding&  $O(M(K_{e}^{2} + K^{2} _{CS}))$&  Increased& No
\\ \hline  
      Li et al. \cite{pr2}   &  Maximizing SE via soft pilot and frequency reuse&    Downlink Multicell M-MIMO &  Uncorrelated Rayleigh fading&  ZF precoding&  NA&  Increased& No
\\ \hline  
    Shahabi et al. \cite{aoa1} &  Decrease complexity with only a slight SE variance&    Uplink multicell M-MIMO &  Multi-path channel model&  Angle of arrival based combining&  $O\left(K^{3.5L} \log\left(\frac{K^L}{\epsilon}\right)\right)$& Not addressed & Not addressed
\\ \hline  
     Omid  et al. \cite{aoa2}  &  Suboptimal pilot allocation via reinforcement learning &    Uplink multicell M-MIMO &  Correlated Rayleigh fading &  Direction of arrival based combining&  NA  & Not addressed & Not addressed
\\ \hline

 \multicolumn{8}{|c|}{\textbf{Mitigating Pilot Contamination with Advanced Signal Processing}}

\\ \hline  
       Lago et al. \cite{si1} &  Improved channel estimation via superimposed pilots&    Uplink multicell M-MIMO &  Correlated Rayleigh fading&  MR, RZF combining&  NA& Not addressed & Not addressed
\\ \hline  
     Shafin et al.  \cite{si2}   &  Improved network performance through superimposed pilots&    Uplink-Downlink multicell M-MIMO &  Parametric channel model&  Direction of departure/arrival based precoding&  NA&  Zero& Yes
       
\\ \hline 
      Mishra et al. \cite{rs3}   &  Improve channel estimation through rate splitting&  Downlink single cell M-MIMO &  Correlated Rayleigh fading&  MR and common precoder&  $O(N_0 \max(4, (4K + 1)^2 (9K + 1)), F_{2})$&  Decreased& Yes

\\ \hline
       Mishra et al. \cite{rs1} &  Enhances SE despite imperfect channel estimates&  Downlink cell-free  MTC M-MIMO  &  Uncorrelated Rayleigh fading&  Heuristic common precoder&  NA&  Decreased& Yes
\\\hline
       Zheng et al. \cite{rs2}&  Improve pilot orthogonality in asynchronous reception&  Downlink cell-free M-MIMO &  Correlated Rayleigh fading&  MMSE, MR combining&  NA& Not addressed & Not addressed
\\ \hline

 \multicolumn{8}{|c|}{\textbf{Pilot Decontamination using Advanced Channel Estimation Methods}}

\\ \hline  
      Xu et al. \cite{dl1}   &  Improve MSE through learning-based pilot design&   Uplink single cell M-MIMO &  Uncorrelated Rayleigh fading&  NA&  $O({KMn_1} + {K{\tau} {n_{L-1}}} +{ \sum_{l=2}^{L-1} n_{l-1}n_l)}$&  Decreased& Yes
\\ \hline  
       Hirose et al. \cite{dl2} &  Improve NMSE of channel through deep learning&     Uplink multicell M-MIMO&Correlated Rayleigh fading&  NA&  NA& Not addressed & Not addressed
\\ \hline  
       Balevi et al.  \cite{dl3}   &  Enhance NMSE of channel via untrained deep learning&     Uplink multicell M-MIMO&  Kronecker channel model&  NA&  $O(M^2)$& Not addressed & Not addressed
\\ \hline  
      Lim et al.  \cite{dl4} &  Deep residual learning based pilot design and channel estimator&     Uplink multicell M-MIMO&  Rayleigh and Nakagami-m channel model&  NA&  $O(N_{ep} N_T N K \times \sum_{m=1}^{M} n_{m-1}n_m F^{2}_{N, m})$& Not addressed & Not addressed
\\ \hline 
    \end{tabular}
    
    \label{tab:my_label}
    \vspace{-0.5cm}
\end{table*}
\end{adjustwidth}

   \vspace{-.5cm}
\section{ Discussion and Relative Merits }
Pilot contamination is a recognized barrier to improving SE in M-MIMO systems. Table 1 provides a comparative analysis of schemes based on major contribution, system considered, channel assumption, combining/precoding scheme, computational complexity, pilot overhead, and scalability. The major contributions of the proposed schemes include enhancing throughput through intelligent pilot assignment or reuse \cite{spa1}, reducing pilot overhead\cite{si1}, and improving channel estimation via advanced techniques \cite{dl1}. Moreover, the proposed schemes are evaluated in two scenarios: cell-free massive MIMO, utilizing distributed antennas without dedicated cell boundaries, and multicell massive MIMO, which employs traditional cell-based structures with coordinated interference control. Furthermore, most papers consider correlated and uncorrelated Rayleigh fading channel assumptions. The key difference is that uncorrelated channels assume an equal signal distribution in all directions, which is impractical, while correlated channel assumptions suggest variations in signal presence among different directions.

\par Different schemes adopt specific combining/precoding techniques based on their complexity and accuracy requirements. Precoding schemes play crucial roles in signal transmission optimization. Match filter (MF), maximum ratio (MR), and zero-forcing (ZF) methods are foundational techniques. Additionally, regularized zero-forcing (RZF) and minimum mean squared error (MMSE) precoding schemes offer advanced capabilities for improved system performance but exhibit higher complexity than conventional schemes \cite{intro2}. Generally, channel estimation requires a minimum of $\tau_p$ pilot symbols, with $\tau_p\ge K$, often reused in neighboring cells. We categorize pilot overhead as 'increased' for schemes that introduce pilot sequence orthogonality between users of neighboring cells, $\tau_p>K$, 'decreased' for schemes where $\tau_p<K$, fewer pilot symbols than $K$ while accommodating all users, and 'not addressed' means a typical pilot reuse scenario. Lastly, the scalability criterion assesses the scheme's ability to accommodate a growing number of devices in a cell while considering the limitation imposed by the coherence block length constraint.

\par Smart pilot assignment schemes \cite{spa1}, \cite{spa2} offer simplicity and efficiency but come with high computation costs. Moreover, graph coloring schemes \cite{gc1}, \cite{gc3} minimize the use of orthogonal pilot signals, yet their effectiveness heavily relies on interference graph criteria. Additionally, soft pilot reuse \cite{pr1} efficiently assigns orthogonal pilot sequences between cells' center and edge users, enhancing performance, but faces challenges due to high pilot signal overhead. Furthermore, the angle of arrival \cite{aoa1} considers the signal's direction as a criterion to distinguish pilot signals of multiple users, but in multi-path scenarios, distinguishing between multiple user angles becomes challenging. Overall, in the ongoing efforts to combat pilot contamination within wireless communication systems, pilot assignment strategies have emerged as pioneering solutions and ongoing research is dedicated to refining their effectiveness \cite{gc4}.

Additionally, introducing the superimposed pilot schemes \cite{si2}, \cite{si1} has addressed the issue of pilot signal length by transmitting pilots alongside data, significantly mitigating pilot contamination. Furthermore, RSMA \cite{rs1}, \cite{rs2} has effectively reduced interference by segregating pilot signals into common and private components, allowing users to decode their data with minimal interference from other users' pilots. Overall, superimposed pilots and RSMA significantly reduce pilot overhead, ultimately increasing spectral efficiency. However, both schemes require advanced signal processing techniques, consequently necessitating high-end hardware support \cite{rs1}.

Furthermore, the increasing utilization of advanced machine learning techniques like DL and CNN is playing a growing role in enhancing pilot design and channel estimation accuracy. A DL-based pilot design approach \cite{dl1} showed promise in optimizing power allocation, but concerns persist regarding its reliance on unsupervised learning. Another  \cite{dl2} introduced NN and CNN-based channel estimation techniques, with CNN demonstrating higher accuracy but potential scalability issues. Additionally, a low-complexity DNN-based channel estimation method \cite{dl3} was presented, with concerns about computational complexity. Lastly, a DL-based approach \cite{dl4}  to mitigate pilot contamination was proposed, showing superior performance even in the presence of hardware impairments, highlighting its interference mitigation potential without prior knowledge. These combined advancements emphasize the dynamic evolution of pilot contamination mitigation strategies, with a clear focus on optimizing SE and efficient resource utilization.

\section{Possible  Future Research Areas }
\par Future research into resource-efficient pilot schemes, such as time-multiplexed pilots \cite{spa1}, \cite{gc3}, superimposed pilots \cite{si1}, and rate splitting \cite{rs1}, could continue evolving due to their potential to effectively mitigate interference. These schemes offer opportunities to optimize the allocation of resources for both pilot and data transmission, thereby reducing overhead and maximizing throughput. Further exploration of these techniques enables researchers to delve into the integration of the aforementioned schemes and fine-tuning resource management for optimal efficiency, thereby improving SE.

\par The increasing demand for wireless connectivity, coupled with the need for a growing number of orthogonal pilot sequences for channel estimation, presents a challenge due to the coherence block length constraint \cite{gc3}. In forthcoming research, it is imperative to prioritize the advancement of intelligent user scheduling schemes \cite{gc4} to effectively manage the projected exponential growth in the number of devices in the coming years. These intelligent scheduling schemes will reduce pilot overhead, address pilot contamination, and enable scalability, ultimately improving the system's throughput. 

\par In the future, incorporating reinforcement learning \cite{aoa2} into pilot assignments entails the creation of an adaptive framework where an agent adjusts pilot assignments based on real-time network conditions, including mobility, interference, and channel quality. The agent's actions, guided by a reward function, aim to optimize throughput and minimize interference. Training methods, such as Q-learning, can be utilized to balance exploration and exploitation, enabling effective pilot assignments.

\par Furthermore, advanced research should focus on joint pilot design and channel estimation techniques \cite{dl4}, using specialized deep neural network architectures and optimization algorithms to concurrently optimize pilot sequences and estimate channels in dynamic wireless environments. This research would create real-time, adaptable solutions seamlessly integrating pilot optimization and channel estimation to address dynamic fading channels and interference challenges.

\section{ Conclusions }
M-MIMO is crucial for 5G and beyond technologies and offers increased SE but relies on accurate channel information, while pilot contamination remains a significant challenge. Pilot contamination arises from short coherence intervals in dynamic wireless environments, causing interference to/from neighboring cells and reducing SE. This paper briefly discusses various mitigation strategies, such as intelligent pilot allocation schemes, interference mitigation schemes, and advanced techniques like deep learning, with the aim of reducing interference and enhancing SE. Future research may explore adaptive pilot assignment with reinforcement learning, resource-efficient pilot schemes, joint pilot design, and channel estimation to optimize SE and adapt to dynamic wireless conditions.


\begin{thebibliography}{00}
\bibitem{intro1} 
L. Sanguinetti, E. Björnson and J. Hoydis, "Toward M-MIMO 2.0: Understanding Spatial Correlation, Interference Suppression, and Pilot Contamination," in IEEE Transactions on Communications, vol. 68, no. 1, pp. 232-257, Jan. 2020, doi: 10.1109/TCOMM.2019.2945792.

\bibitem{intro2} 
O. Elijah, C. Y. Leow, T. A. Rahman, S. Nunoo and S. Z. Iliya, "A Comprehensive Survey of Pilot Contamination in M-MIMO—5G System," in IEEE Communications Surveys and Tutorials, vol. 18, no. 2, pp. 905-923, Secondquarter 2016, doi: 10.1109/COMST.2015.2504379.

\bibitem{spa1} 
X. Zhu, Z. Wang, L. Dai and C. Qian, "Smart Pilot Assignment for M-MIMO," in IEEE Communications Letters, vol. 19, no. 9, pp. 1644-1647, Sept. 2015, doi: 10.1109/LCOMM.2015.2409176.

\bibitem{spa2} 
T. H. Nguyen, T. V. Chien, H. Q. Ngo, X. N. Tran and E. Björnson, "Pilot Assignment for Joint Uplink-Downlink SE Enhancement in M-MIMO Systems With Spatial Correlation," in IEEE Transactions on Vehicular Technology, vol. 70, no. 8, Aug. 2021, doi: 10.1109/TVT.2021.3091020.

\bibitem{gc1} 
H. Liu, J. Zhang, S. Jin and B. Ai, "Graph Coloring Based Pilot Assignment for Cell-Free M-MIMO Systems," in IEEE Transactions on Vehicular Technology, vol. 69, no. 8, pp. 9180-9184, Aug. 2020, doi: 10.1109/TVT.2020.3000496.

\bibitem{gc2}
W. Zeng, Y. He, B. Li and S. Wang, "Pilot Assignment for Cell-Free M-MIMO Systems Using a Weighted Graphic Framework," in IEEE Transactions on Vehicular Technology, vol. 70, no. 6, pp. 6190-6194, June 2021, doi: 10.1109/TVT.2021.3076440.


\bibitem{gc3}
M. K. Saeed and A. Khokhar, ‘Smart Pilot Assignment for IoT in Massive MIMO Systems: A Path Towards Scalable IoT Infrastructure’, in IEEE ICC (International Conference on Communication) 2024, June 2024
https://doi.org/10.48550/arXiv.2404.10188
 arXiv:2404. 10188, 2024.

\bibitem{gc4}
M. K. Saeed, A. E. Kamal, and A. Khokhar, ‘Mitigating Pilot Contamination and Enabling IoT Scalability in Massive MIMO Systems’, in GLOBECOM 2023-2023 IEEE Global Communications Conference, 2023, pp. 1620–1625.

\bibitem{pr1}

X. Zhu et al., "Soft Pilot Reuse and Multicell Block Diagonalization Precoding for M-MIMO Systems," in IEEE Transactions on Vehicular Technology, vol. 65, no. 5, pp. 3285-3298, May 2016, doi: 10.1109/TVT.2015.2445795.

\bibitem{pr2}

Y. Li, R. Wang and Z. Zhang, "M-MIMO Downlink Goodput Analysis With Soft Pilot or Frequency Reuse," in IEEE Wireless Communications Letters, vol. 7, no. 3, pp. 448-451, June 2018, doi: 10.1109/LWC.2017.2783336.




\bibitem{aoa1} 

Shahabi, S. M., Mosleh, M. A., and Ardebilipour, M. (2020). Low-complexity AoA-driven pilot assignment for multi-cell M-MIMO systems. Physical Communication, 42, 101118.

\bibitem{aoa2} 

Y. Omid, S. M. Hosseini, S. M. Shahabi, M. Shikh-Bahaei and A. Nallanathan, "AoA-Based Pilot Assignment in M-MIMO Systems Using Deep Reinforcement Learning," in IEEE Communications Letters, vol. 25, no. 9, pp. 2948-2952, Sept. 2021, doi: 10.1109/LCOMM.2021.3089234.


\bibitem{si1} 
L. A. Lago, Y. Zhang, N. Akbar, Z. Fei, N. Yang and Z. He, "Pilot Decontamination Based on Superimposed Pilots Assisted by Time-Multiplexed Pilots in M-MIMO Networks," in IEEE Transactions on Vehicular Technology, vol. 69, no. 1, pp. 405-417, Jan. 2020, doi: 10.1109/TVT.2019.2949605.

\bibitem{si2}
R. Shafin and L. Liu, "Superimposed Pilot for Multi-Cell Multi-User Massive FD-MIMO Systems," in IEEE Transactions on Wireless Communications, vol. 19, no. 5, pp. 3591-3606, May 2020, doi: 10.1109/TWC.2020.2975551.






\bibitem{rs3}
A. Mishra, Y. Mao, C. K. Thomas, L. Sanguinetti and B. Clerckx, "Mitigating Intra-Cell Pilot Contamination in M-MIMO: A Rate Splitting Approach," in IEEE Transactions on Wireless Communications, vol. 22, no. 5, pp. 3472-3487, May 2023, doi: 10.1109/TWC.2022.3218897.

\bibitem{rs1} 
 
A. Mishra, Y. Mao, L. Sanguinetti and B. Clerckx, "Rate-Splitting Assisted Massive Machine-Type Communications in Cell-Free M-MIMO," in IEEE Communications Letters, vol. 26, no. 6, pp. 1358-1362, June 2022, doi: 10.1109/LCOMM.2022.3160511.



\bibitem{rs2} 
J. Zheng, J. Zhang, J. Cheng, V. C. M. Leung, D. W. K. Ng and B. Ai, "Asynchronous Cell-Free M-MIMO With Rate-Splitting," in IEEE Journal on Selected Areas in Communications, vol. 41, no. 5, pp. 1366-1382, May 2023, doi: 10.1109/JSAC.2023.3240709.


\bibitem{dl1}
J. Xu, P. Zhu, J. Li and X. You, "Deep Learning-Based Pilot Design for Multi-User Distributed M-MIMO Systems," in IEEE Wireless Communications Letters, vol. 8, no. 4, pp. 1016-1019, Aug. 2019, doi: 10.1109/LWC.2019.2904229.

\bibitem{dl2}
H. Hirose, T. Ohtsuki and G. Gui, "Deep Learning-Based Channel Estimation for M-MIMO Systems With Pilot Contamination," in IEEE Open Journal of Vehicular Technology, vol. 2, pp. 67-77, 2021, doi: 10.1109/OJVT.2020.3045470.

\bibitem{dl3}
E. Balevi, A. Doshi and J. G. Andrews, "M-MIMO Channel Estimation With an Untrained Deep Neural Network," in IEEE Transactions on Wireless Communications, vol. 19, no. 3, pp. 2079-2090, March 2020, doi: 10.1109/TWC.2019.2962474.
\bibitem{dl4}
B. Lim, W. J. Yun, J. Kim and Y. -C. Ko, "Joint Pilot Design and Channel Estimation Using Deep Residual Learning for Multi-Cell M-MIMO Under Hardware Impairments," in IEEE Transactions on Vehicular Technology, vol. 71, no. 7, pp. 7599-7612, July 2022, doi: 10.1109/TVT.2022.3170556.




\end{thebibliography}
\end{document}